\documentclass[%
reprint,
showpacs,preprintnumbers,
nofootinbib,
amsmath,amssymb,
aps,
prd,
floatfix,
]{revtex4-1}

\usepackage{graphicx}
\usepackage{dcolumn}
\usepackage{bm}

\usepackage{url}

\newcommand{\boss}[2]{\ensuremath{\rlap{\kern-2.5pt\ensuremath{\overset{\scriptscriptstyle(-)}{\phantom{#1}}}}{\ensuremath{{#1}_{#2}}}}}

\begin{document}

\preprint{\begin{tabular}{l}
\texttt{EURONU-WP6-11-43}
\\
\texttt{arXiv:1109.4033 [hep-ph]}
\end{tabular}}

\title{Status of 3+1 Neutrino Mixing}

\author{Carlo Giunti}
\email{giunti@to.infn.it}
\altaffiliation[also at ]{Department of Theoretical Physics, University of Torino, Italy}
\affiliation{INFN, Sezione di Torino, Via P. Giuria 1, I--10125 Torino, Italy}

\author{Marco Laveder}
\email{laveder@pd.infn.it}
\affiliation{Dipartimento di Fisica ``G. Galilei'', Universit\`a di Padova,
and
INFN, Sezione di Padova,
Via F. Marzolo 8, I--35131 Padova, Italy}

\date{\today}

\begin{abstract}
We present an update of our analysis of short-baseline neutrino oscillation data
in the framework of 3+1 neutrino mixing
taking into account the recent update of MiniBooNE antineutrino data
and the recent results of the MINOS search for
$\nu_{\mu}$ disappearance into sterile neutrinos
(the more complicated 3+2 neutrino mixing is not needed since
the
CP-violating difference between
MiniBooNE neutrino and antineutrino data has diminished).
The results of our fits of short-baseline neutrino oscillation
data including the MiniBooNE low-energy anomaly
(now present both in the neutrino and antineutrino data)
leads to a strong tension between appearance and disappearance data.
Hence,
it seems likely that the low-energy anomaly
is not due to
$\boss{\nu}{\mu}\to\boss{\nu}{e}$
transitions.
Excluding the MiniBooNE low-energy anomaly,
appearance and disappearance data are marginally compatible.
The global analysis has the best-fit point at
$\Delta{m}^2_{41} \approx 5.6\,\text{eV}^2$,
which is rather large in comparison with cosmological bounds,
but there are three regions within $1\sigma$
at
$
\Delta{m}^2_{41}
\approx
1.6
\,,\,
1.2
\,,\,
0.91
\,
\text{eV}^2
$.
We also show that
the data on the Gallium neutrino anomaly
favor values of
$\Delta{m}^2_{41}$ larger than about $1 \, \text{eV}^2$.
\end{abstract}


\maketitle

\section{Introduction}
\label{Introduction}

The possible existence of sterile neutrinos is an exciting possibility
which could open a powerful window on our view of the physics beyond the Standard Model.
The short-baseline (SBL) neutrino oscillation experiment
LSND \cite{hep-ex/0104049}
discovered in the late 90's a signal which can be due to
$\bar\nu_{\mu}\to\bar\nu_{e}$ oscillations
generated by a neutrino squared-mass splitting
$\Delta{m}^2_{\text{SBL}}$
of the order of $1 \, \text{eV}^2$,
which is much larger than the
well-established solar (SOL) and atmospheric (ATM) squared-mass splittings,
$
\Delta{m}^2_{\text{SOL}}
=
(7.6 \pm 0.2) \times 10^{-5} \, \text{eV}^2
$
\cite{1010.0118}
and
$
\Delta{m}^2_{\text{ATM}}
=
2.32 {}^{+0.12}_{-0.08} \times 10^{-3} \, \text{eV}^2
$
\cite{1103.0340}.
In order to have more than two independent squared-mass splittings
the number of massive neutrinos must be larger than three.
In this case the flavor neutrino basis is composed by the three
known active neutrinos
$\nu_{e}$,
$\nu_{\mu}$,
$\nu_{\tau}$
and by one or more sterile neutrinos
$\nu_{s1}$,
$\nu_{s2}$,
\ldots,
which do not have standard weak interactions and do not contribute
to the number of active neutrinos determined by LEP experiments
through the measurement of the invisible width of the $Z$ boson,
$N_{a} = 2.9840 \pm 0.0082$
\cite{hep-ex/0509008}.

Schemes of neutrino mixing with sterile neutrinos have been studied by several authors
(see Refs.~\cite{hep-ph/9812360,hep-ph/0405172,hep-ph/0606054,GonzalezGarcia:2007ib}),
with more attention to the simple schemes with
one or two sterile neutrinos
(four- and five-neutrino mixing, respectively).
Since the three active neutrinos must have large mixing with the three massive neutrinos
which generate
$\Delta{m}^2_{\text{SOL}}$
and
$\Delta{m}^2_{\text{ATM}}$
and no effect of sterile neutrinos has been seen in solar and atmospheric neutrino data,
the mixing schemes with sterile neutrinos
must be perturbations of the standard three-neutrino mixing scheme
in which the three active neutrinos
$\nu_{e}$,
$\nu_{\mu}$,
$\nu_{\tau}$
are superpositions of three massive neutrinos
$\nu_1$,
$\nu_2$,
$\nu_3$
with respective masses
$m_1$,
$m_2$,
$m_3$,
such that
$
\Delta{m}^2_{\text{SOL}}
=
\Delta{m}^2_{21}
$
and
$
\Delta{m}^2_{\text{ATM}}
=
|\Delta{m}^2_{31}|
\simeq
|\Delta{m}^2_{32}|
$,
with
$\Delta{m}^2_{kj}=m_k^2-m_j^2$.
Moreover,
standard analyses of the
Cosmic Microwave Background and Large-Scale Structures data
constrain the neutrino masses in the case of three-neutrino mixing
to be much smaller than 1 eV
\cite{0805.2517,0910.0008,0911.5291,1006.3795}
and
are compatible with the existence of one or two sterile neutrinos
which have been thermalized in the early Universe
\cite{1109.2767}
only if the masses of the additional, mainly sterile, massive neutrinos are smaller than about 1 eV
\cite{1006.5276,1102.4774,1104.0704,1104.2333,1106.5052,1108.4136}.
Also
Big-Bang Nucleosynthesis data
are compatible with the existence of sterile neutrinos
which have been thermalized in the early Universe
\cite{astro-ph/0408033,1001.4440},
with the indication however that schemes with more than one sterile neutrino are disfavored
\cite{1103.1261,1108.4136}.
Hence,
the schemes with sterile neutrinos which are currently under consideration are the 3+1 and 3+2 schemes
in which
$\nu_{e}$,
$\nu_{\mu}$,
$\nu_{\tau}$
are mainly mixed with
$\nu_1$,
$\nu_2$,
$\nu_3$, whose masses are much smaller than 1 eV
and there are one or two additional massive neutrinos,
$\nu_4$ and
$\nu_5$,
which are mainly sterile
and have masses of the order of 1 eV.
Short-baseline oscillations corresponding to the
LSND
$\bar\nu_{\mu}\to\bar\nu_{e}$
signal
are generated by the large squared-mass differences
$\Delta{m}^2_{41}$
and
$\Delta{m}^2_{51}$.

The LSND signal was not seen in the
KARMEN $\bar\nu_{\mu}\to\bar\nu_{e}$
\cite{hep-ex/0203021},
NOMAD $\nu_{\mu}\to\nu_{e}$
\cite{hep-ex/0306037}
and
MiniBooNE $\nu_{\mu}\to\nu_{e}$
\cite{0812.2243}
short-baseline experiments.
However,
in July 2010 the interest in the LSND signal has been revived by the observation of
a compatible signal in the
MiniBooNE $\bar\nu_{\mu}\to\bar\nu_{e}$
short-baseline experiment
\cite{1007.1150}.
The MiniBooNE and LSND antineutrino data have been analyzed in several papers
in conjunction with the data of other short-baseline experiments
in the framework of 3+1 and 3+2 schemes
\cite{1007.4171,1012.0267,1103.4570,1107.1452}.
In this paper we update the
3+1 analysis presented in Ref.~\cite{1107.1452}
by taking into account the update of MiniBooNE antineutrino data
presented in Refs.~\cite{1111.1375,Djurcic-NUFACT2011}
and the recent results of the MINOS search for
$\nu_{\mu}$ disappearance into sterile neutrinos
\cite{1104.3922}.

In 3+1 schemes we have the squared-mass hierarchy
\begin{equation}
\Delta{m}^2_{21}
\ll
\Delta{m}^2_{31}
\ll
\Delta{m}^2_{41}
\,,
\label{hierarchy}
\end{equation}
and we consider four-neutrino mixing as a perturbation of three-neutrino mixing:
\begin{equation}
|U_{e4}|^2
\,,\,
|U_{\mu4}|^2
\,,\,
|U_{\tau4}|^2
\,,\,
\ll 1
\,,
\quad
|U_{s4}|^2 \simeq 1
\,.
\label{perturbation}
\end{equation}
The
effective flavor transition and survival probabilities
in short-baseline experiments
are given by
\begin{align}
\null & \null
P_{\boss{\nu}{\alpha}\to\boss{\nu}{\beta}}^{\text{SBL}}
=
\sin^{2} 2\vartheta_{\alpha\beta}
\sin^{2}\left( \frac{\Delta{m}^2_{41} L}{4E} \right)
\qquad
(\alpha\neq\beta)
\,,
\label{trans}
\\
\null & \null
P_{\boss{\nu}{\alpha}\to\boss{\nu}{\alpha}}^{\text{SBL}}
=
1
-
\sin^{2} 2\vartheta_{\alpha\alpha}
\sin^{2}\left( \frac{\Delta{m}^2_{41} L}{4E} \right)
\,,
\label{survi}
\end{align}
for
$\alpha,\beta=e,\mu,\tau,s$,
with
\begin{align}
\null & \null
\sin^{2} 2\vartheta_{\alpha\beta}
=
4 |U_{\alpha4}|^2 |U_{\beta4}|^2
\,,
\label{transsin}
\\
\null & \null
\sin^{2} 2\vartheta_{\alpha\alpha}
=
4 |U_{\alpha4}|^2 \left( 1 - |U_{\alpha4}|^2 \right)
\,.
\label{survisin}
\end{align}

In this paper we do not consider 3+2 neutrino mixing,
because the new MiniBooNE antineutrino data
\cite{1111.1375,Djurcic-NUFACT2011}
do not show a sufficient difference from the
MiniBooNE neutrino data \cite{0812.2243}
to motivate the consideration of the much more complicated 3+2 neutrino mixing
which could explain a difference through
CP violation in short-baseline oscillations
\cite{hep-ph/0305255,hep-ph/0609177,0906.1997,0705.0107,1007.4171,1103.4570,1107.1452}.
In Ref.~\cite{1107.1452}
we have shown that such an effect would be the main motivation for
preferring 3+2 mixing over 3+1 mixing.

The plan of the paper is as follows.
In Sections~\ref{MiniBooNE} and \ref{MINOS}
we describe, respectively,
our analysis
of the new MiniBooNE antineutrino data
and
that of MINOS data on the search for
$\nu_{\mu}$ disappearance into sterile neutrinos.
In Section~\ref{Global}
we present the results of the global fit of
short-baseline oscillation data
and in Section~\ref{Conclusions}
we draw our conclusions.

\section{MiniBooNE}
\label{MiniBooNE}

\begin{figure}[t!]
\includegraphics*[width=\linewidth]{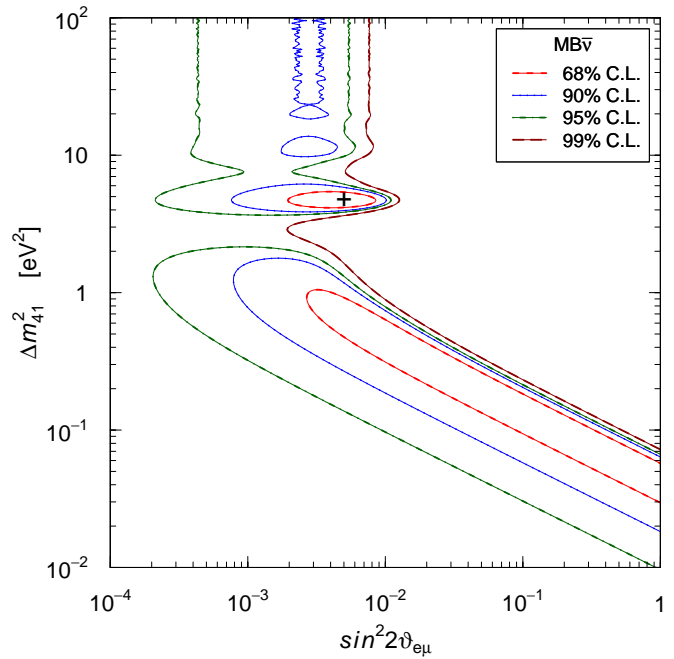}
\caption{ \label{mba-low}
Allowed regions in the
$\sin^{2}2\vartheta_{e\mu}$--$\Delta{m}^2_{41}$ plane
obtained from the fit of MiniBooNE antineutrino data
\protect\cite{1111.1375,Djurcic-NUFACT2011},
including the low-energy bins
from
$200 \, \text{MeV}$
to
$475 \, \text{MeV}$.
The best-fit point at
$\sin^{2}2\vartheta_{e\mu} = 0.005$
and
$\Delta{m}^2_{41} = 4.68 \, \text{eV}^2$
is indicated by a cross.
}
\end{figure}

The MiniBooNE collaboration presented recently a preliminary update of their antineutrino data
obtained with
$8.58\times10^{20}$
Protons on Target (PoT)
\cite{1111.1375,Djurcic-NUFACT2011},
which increases the amount of data by a factor of about 3/2 with respect to the data published in Ref.~\cite{1007.1150},
which were obtained with
$5.66\times10^{20}$ PoT.
The new data show two interesting new features:
\begin{enumerate}
\item
The antineutrino data have an anomalous low-energy excess
similar to that of the neutrino data
\cite{0812.2243}.
\item
The $\bar\nu_{\mu}\to\bar\nu_{e}$
signal in the three energy bins from
$475 \, \text{MeV}$
to
$800 \, \text{MeV}$
has slightly diminished
with respect to that published in Ref.~\cite{1007.1150}.
\end{enumerate}

The first new feature raises the interesting question if
the MiniBooNE low-energy anomaly can be fitted by oscillations.
In order to answer to this question we consider the fit of MiniBooNE
data with and without the three low-energy bins
from
$200 \, \text{MeV}$
to
$475 \, \text{MeV}$.
The second new feature may be the consequence of the fluctuations of the signal
around the true value.
A consequence of the new data is that the difference between MiniBooNE neutrino and antineutrino data
has diminished,
lessening the need of CP violation that was suggested by the previous data.
Hence,
there are less motivations for considering
3+2 neutrino mixing
\cite{hep-ph/0305255,hep-ph/0609177,0906.1997,0705.0107,1007.4171,1103.4570,1107.1452},
in which
$\nu_{\mu}\to\nu_{e}$
and
$\bar\nu_{\mu}\to\bar\nu_{e}$
oscillations can be different
if
$\text{Im}[U_{e4}^{*}U_{\mu4}U_{e5}U_{\mu5}^{*}] \neq 0$.
Moreover,
as we argued in Ref.~\cite{1107.1452},
3+1 neutrino mixing is preferable on 3+2 mixing
for its simplicity and for the natural correspondence of one new entity (a sterile neutrino) with a new effect (short-baseline oscillations).
In Ref.~\cite{1107.1452} we have also shown
that
the improvement of the parameter goodness of fit in 3+2 schemes with respect to 3+1 schemes
is mainly a statistical effect due to an increase of the number of parameters,
in agreement with the results of Ref.~\cite{1103.4570}.
Therefore,
in this paper we consider only 3+1 neutrino mixing.

We analyzed the MiniBooNE antineutrino data
extracted from the figures presented in Refs.~\cite{1111.1375,Djurcic-NUFACT2011}.
We followed the method described in the MiniBooNE web page of the data release relative to Ref.~\cite{1007.1150},
modified by rescaling the predicted signal by the ratio of POT.
Furthermore,
in order to reproduce the allowed regions in the
$\sin^22\vartheta$--$\Delta{m}^2$
plane presented in Refs.~\cite{1111.1375,Djurcic-NUFACT2011},
we increased the background uncertainty by a factor 1.1.
In the fit we considered also
the $\bar\nu_{\mu}$ data obtained with $5.66 \times 10^{20}$ POT \cite{1007.1150},
which are important because of the correlated uncertainties
of $\bar\nu_{e}$ and $\bar\nu_{\mu}$ data.

\begin{figure}[t!]
\includegraphics*[width=\linewidth]{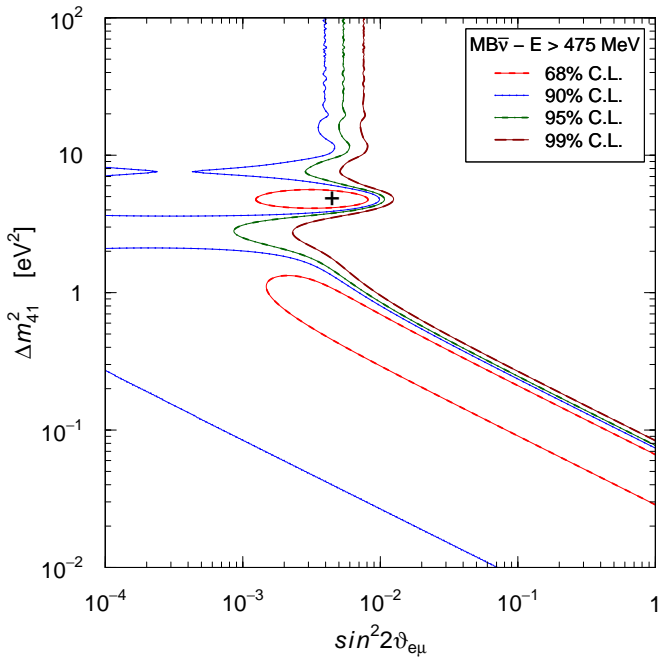}
\caption{ \label{mba-hig}
Allowed regions in the
$\sin^{2}2\vartheta_{e\mu}$--$\Delta{m}^2_{41}$ plane
obtained from the fit of MiniBooNE antineutrino data
with energy $E>475 \, \text{MeV}$
\protect\cite{1111.1375,Djurcic-NUFACT2011}.
The best-fit point at
$\sin^{2}2\vartheta_{e\mu} = 0.0045$
and
$\Delta{m}^2_{41} = 4.79 \, \text{eV}^2$
is indicated by a cross.
}
\end{figure}

The results of our fits are shown in Figs.~\ref{mba-low} and \ref{mba-hig},
respectively,
for all MiniBooNE antineutrino energy bins and
for the energy bins above $475 \, \text{MeV}$.
One can see that
the allowed regions are similar to the corresponding ones presented in
Refs.~\cite{1111.1375,Djurcic-NUFACT2011}.

\begin{figure}[t!]
\includegraphics*[width=\linewidth]{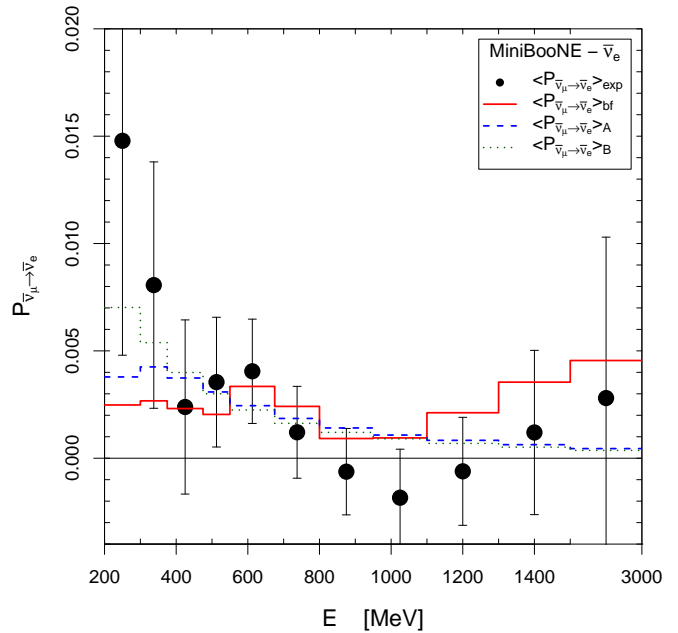}
\caption{ \label{mba-low-hst-pme}
Fit of MiniBooNE antineutrino data
\protect\cite{1111.1375,Djurcic-NUFACT2011}
(points with error bars).
The red solid line corresponds to the best fit
($\sin^{2}2\vartheta_{e\mu} = 0.005$
and
$\Delta{m}^2_{41} = 4.68 \, \text{eV}^2$).
The blue dashed and green dotted lines correspond,
respectively,
to:
A:
$\sin^{2}2\vartheta_{e\mu} = 0.005$
and
$\Delta{m}^2_{41} = 0.8 \, \text{eV}^2$;
B:
$\sin^{2}2\vartheta_{e\mu} = 0.01$
and
$\Delta{m}^2_{41} = 0.5 \, \text{eV}^2$.
}
\end{figure}

As one can see from Fig.~\ref{mba-low-hst-pme},
the low-energy anomaly is not well fitted for the best-fit value of the
oscillation parameters,
$\sin^{2}2\vartheta_{e\mu} = 0.005$
and
$\Delta{m}^2_{41} = 4.68 \, \text{eV}^2$,
but can be fitted in the case of a lower value of $\Delta{m}^2_{41}$.
In the example A
($\sin^{2}2\vartheta_{e\mu} = 0.005$
and
$\Delta{m}^2_{41} = 0.8 \, \text{eV}^2$,
which is within the 68\% C.L. allowed region in Fig.~\ref{mba-low})
the value of $\sin^{2}2\vartheta_{e\mu}$ is the same as in the best-fit point
but the lower value of $\Delta{m}^2_{41}$
increases
$\langle P_{\bar\nu_{\mu}\to\bar\nu_{e}} \rangle$
in the low-energy bins.
In the example B
($\sin^{2}2\vartheta_{e\mu} = 0.01$
and
$\Delta{m}^2_{41} = 0.5 \, \text{eV}^2$,
which is also within the 68\% C.L. allowed region in Fig.~\ref{mba-low})
the larger value of $\sin^{2}2\vartheta_{e\mu}$
and
the smaller value of $\Delta{m}^2_{41}$
allow us to fit the low-energy anomaly even better.
Although by eye the lines corresponding to the cases A and B
may appear to fit all the MiniBooNE antineutrino data better than the best-fit line,
the best-fit point has a lower value of $\chi^2$
because of the correlations of the uncertainties of the bins,
which are given by the covariance matrix of the MiniBooNE data release relative to Ref.~\cite{1007.1150}.

\section{MINOS}
\label{MINOS}

The MINOS collaboration presented recently
\cite{1104.3922}
the updated results of a search for
$\nu_{\mu}$ disappearance into sterile neutrinos
obtained by comparing the samples of neutral current (NC) events
measured at the
near detector (ND)
and
far detector (FD).
In the MINOS experiment
the neutrino beam is produced through the decay of pions generated by 120 GeV protons hitting a graphite target.
The pions fly in a 675 m long decay pipe.
The near and far detectors are located, respectively,
at the distances
$L_{\text{ND}} = 1.04 \, \text{km}$
and
$L_{\text{FD}} = 735 \, \text{km}$
from the target.
The analysis presented in Ref.~\cite{1104.3922}
limits
$|U_{\mu4}|^2$ below 0.019 at 90\% C.L.
assuming that there are no oscillations before the near detector
and that the oscillations are completely averaged in the far detector.
Since the neutrino energy range goes from about 1 GeV to about 20 GeV,
the first condition is satisfied for
$\Delta{m}^2_{41} \lesssim 1 \, \text{eV}^2$
and the second condition is satisfied for
$\Delta{m}^2_{41} \gtrsim 0.2 \, \text{eV}^2$.
Hence, the range of $\Delta{m}^2_{41}$ for which the bound on
$|U_{\mu4}|^2$
presented in Ref.~\cite{1104.3922} is limited.

\begin{figure}[t!]
\includegraphics*[width=\linewidth]{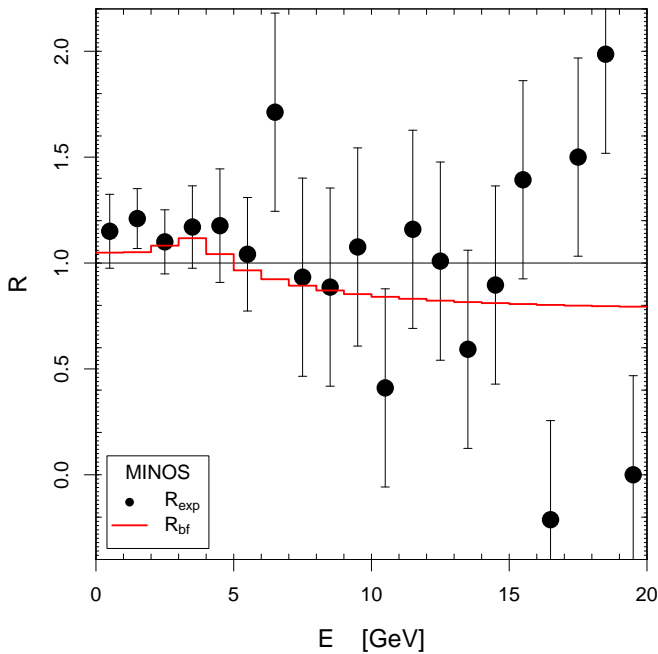}
\caption{ \label{hst-mns}
Fit of MINOS data extracted from Ref.~\protect\cite{1104.3922}.
The red solid line corresponds to the best fit
($\sin^{2}2\vartheta_{e\mu} = 0.005$
and
$\Delta{m}^2_{41} = 4.68 \, \text{eV}^2$).
}
\end{figure}

Since we consider higher values of
$\Delta{m}^2_{41}$,
we analyzed the data presented in Ref.~\cite{1104.3922}
taking into account possible oscillations before the near detector.
From our extraction of the data in Ref.~\cite{1104.3922}
we obtained the values shown in Fig.~\ref{hst-mns}
for the ratio
\begin{equation}
R
=
\frac{\langle 1 - P_{\nu_{\mu}\to\nu_{s}} \rangle_{\text{FD}}}{\langle 1 - P_{\nu_{\mu}\to\nu_{s}} \rangle_{\text{ND}}}
\,.
\label{ratio-minos}
\end{equation}
In the fit we considered a fully correlated 2.9\% systematic uncertainty \cite{1104.3922}.
We also took into account as fully correlated the uncertainties
given in Ref.~\cite{1104.3922} on the possible effect of
$\nu_{e}$ appearance in the far detector due to
$|U_{e3}|^2 < 0.040$,
which is the 2010 MINOS 90\% limit \cite{1006.0996}.
This value is compatible with the recent results of the
T2K
\cite{1106.2822}
and
MINOS
\cite{1108.0015}
experiments
which are in favor of $\nu_{e}$ appearance
\cite{1106.6028,1108.1376}.

We calculated the oscillation probability at the near detector
with the approximate method derived in Ref.~\cite{1105.5946},
which takes into account the partial decoherence
of the neutrino state at the production
due to the fact that the decay length of the parent pion and the length of the decay pipe are comparable with the
distance from the target to the near detector.
For completeness,
we took into account also possible oscillations in the far detector
due to $\Delta{m}^2_{41}$,
but we neglected for simplicity possible
$\nu_{\mu}\to\nu_{s}$
transitions due to $\Delta{m}^2_{31}$
(see Eq.~(11) of Ref.~\cite{1001.0336}).
This is equivalent to assuming a negligible value for $|U_{s3}|$.

\begin{figure}[t!]
\includegraphics*[width=\linewidth]{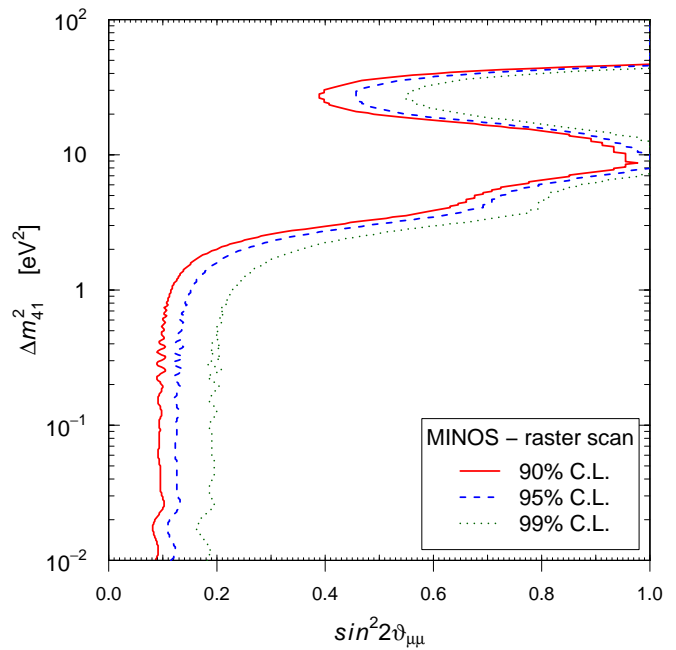}
\caption{ \label{mns-ras-lin}
Raster-scan upper bound on $\sin^{2}2\vartheta_{\mu\mu}$
as a function of
$\Delta{m}^2_{41}$ plane
obtained from the fit of MINOS neutral current data
\protect\cite{1104.3922}.
}
\end{figure}

We averaged the oscillation probabilities over the neutrino flux and the neutral current cross section,
taking into account the energy resolution of the detector:
\begin{align}
\langle P \rangle
=
I^{-1}
\int
\null & \null
\mathrm{d}E_{\text{reco}}
\mathrm{d}E_{\text{h}}
\mathrm{d}E_{\nu}
\mathrm{d}x
\,
\frac{\mathrm{d}^2\sigma_{\text{NC}}}{\mathrm{d}x \mathrm{d}E_{\text{h}}}
\,
\psi_{\text{res}}(E_{\text{h}},E_{\text{reco}})
\nonumber
\\
\null & \null
\times
\Phi_{\nu}(E_{\nu})
\,
P(E_{\nu})
\,,
\label{ave-mns}
\end{align}
where $I$ is the same integral without $P(E_{\nu})$.
We took the neutrino flux $\Phi_{\nu}(E_{\nu})$ as a function of the neutrino energy $E_{\nu}$
from Fig.~2.10 of Ref.~\cite{Loiacono-2010}.
The differential cross section has been approximated by the deep-inelastic cross section
on a isoscalar target
using the NNLO MSTW 2008 set of parton distribution functions
\cite{0901.0002}.
For the energy resolution function
$\psi_{\text{res}}(E_{\text{h}},E_{\text{reco}})$
which connects the hadronic energy $E_{\text{h}}$ to the reconstructed energy
$E_{\text{reco}}$
we used a Gaussian distribution
with standard deviation
$56\% / \sqrt{E_{\text{reco}}}$
\cite{1104.3922}.

Figure~\ref{mns-ras-lin} shows the
upper bound on $\sin^{2}2\vartheta_{\mu\mu}$
as a function of
$\Delta{m}^2_{41}$ plane
obtained with a raster scan,
which can be compared with that obtained by the MINOS collaboration
in Ref.~\cite{1104.3922}.
One can see that for
$\Delta{m}^2_{41} \lesssim 1 \, \text{eV}^2$
we have the limit
$\sin^{2}2\vartheta_{\mu\mu} \lesssim 0.09$
at 90\% C.L.,
with wiggles due to oscillations in the far detector.
Using Eq.~(\ref{survisin}),
this limit corresponds to
$|U_{\mu4}|^2 \lesssim 0.023$,
which is about the same as that
obtained by the MINOS collaboration in Ref.~\cite{1104.3922}.
For $\Delta{m}^2_{41} \gtrsim 1 \, \text{eV}^2$
the upper bound on
$\sin^{2}2\vartheta_{\mu\mu}$ rapidly disappears,
in agreement with the discussion above and that in Ref.~\cite{1105.5946}.

\begin{figure}[t!]
\includegraphics*[width=\linewidth]{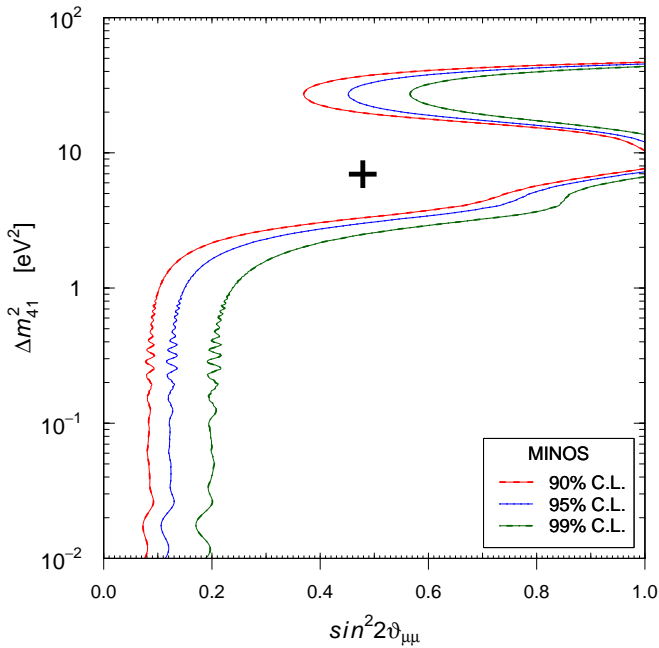}
\caption{ \label{mns-con-lin}
Exclusion curves in the
$\sin^{2}2\vartheta_{\mu\mu}$--$\Delta{m}^2_{41}$ plane
obtained from the fit of MINOS neutral current data
\protect\cite{1104.3922}.
The best-fit point at
$\sin^{2}2\vartheta_{\mu\mu} = 0.48$
and
$\Delta{m}^2_{41} = 6.76 \, \text{eV}^2$
is indicated by a cross.
}
\end{figure}

Figure \ref{mns-con-lin}
shows the exclusion curves in the
$\sin^{2}2\vartheta_{\mu\mu}$--$\Delta{m}^2_{41}$ plane
obtained with a two-parameters least-squares analysis.
Also in this figure
one can see that the oscillations in the near detector weaken the bound on
$|U_{\mu4}|^2$
for
$\Delta{m}^2_{41} \gtrsim 1 \, \text{eV}^2$.
The best-fit point at
$\sin^{2}2\vartheta_{\mu\mu} = 0.48$
and
$\Delta{m}^2_{41} = 6.76 \, \text{eV}^2$
allows to fit the small positive values of $R$ measured in the low-energy bins,
which have the smaller uncertainties.
This effect is in agreement with the discussion in Ref.~\cite{1105.5946},
where it is explained that for a value of
$\Delta{m}^2_{41}$ such that the low-energy bins correspond to the first oscillation minimum at the near detector
the denominator in Eq.~(\ref{ratio-minos})
is smaller than the numerator, leading to $R>1$.

\begin{table*}[t!]
\begin{center}
\begin{tabular}{cccccc}
&
&
LOW
&
LOW+GAL
&
HIG
&
HIG+GAL
\\
\hline
 No Osc. & $\chi^{2}$ & $ 174.8 $ & $ 186.8 $ & $ 157.8 $ & $ 169.8 $ \\
 & NDF & $ 133 $ & $ 137 $ & $ 127 $ & $ 131 $ \\
 & GoF & $ 0.0088 $ & $ 0.003 $ & $ 0.033 $ & $ 0.013 $ \\
\hline 3+1 & $\chi^{2}_{\text{min}}$ & $ 134.9 $ & $ 142.2 $ & $ 120.7 $ & $ 128.0 $ \\
 & NDF & $ 130 $ & $ 134 $ & $ 124 $ & $ 128 $ \\
 & GoF & $ 0.37 $ & $ 0.30 $ & $ 0.57 $ & $ 0.48 $ \\
 & $\Delta{m}^2_{41} [\text{eV}^2]$ & $ 5.6 $ & $ 5.6 $ & $ 5.6 $ & $ 5.6 $ \\
 & $|U_{e4}|^2$ & $ 0.032 $ & $ 0.037 $ & $ 0.033 $ & $ 0.038 $ \\
 & $|U_{\mu4}|^2$ & $ 0.014 $ & $ 0.012 $ & $ 0.013 $ & $ 0.011 $ \\
 & $\sin^22\vartheta_{e\mu}$ & $ 0.0018 $ & $ 0.0018 $ & $ 0.0017 $ & $ 0.0017 $ \\
 & $\sin^22\vartheta_{ee}$ & $ 0.12 $ & $ 0.14 $ & $ 0.13 $ & $ 0.14 $ \\
 & $\sin^22\vartheta_{\mu\mu}$ & $ 0.054 $ & $ 0.049 $ & $ 0.05 $ & $ 0.045 $ \\
\hline PG & $\Delta\chi^{2}_{\text{min}}$ & $ 15.8 $ & $ 15.8 $ & $ 9.3 $ & $ 9.2 $ \\
 & NDF & $ 2 $ & $ 2 $ & $ 2 $ & $ 2 $ \\
 & GoF & $4\times10^{-4} $ & $4\times10^{-4} $ & $ 0.01 $ & $ 0.01 $ \\
\hline
\end{tabular}
\caption{ \label{bef}
Values of
$\chi^{2}$,
number of degrees of freedom (NDF),
goodness-of-fit (GoF)
and
best-fit values
of the 3+1 oscillation parameters
obtained from global fits of the data of short-baseline neutrino oscillation experiments.
The first three lines correspond to the case of no oscillations (No Osc.).
The following nine lines correspond to the case of 3+1 mixing.
The last three lines give the parameter goodness-of-fit (PG) \protect\cite{hep-ph/0304176}
obtained by comparing the global best-fit with the
best fits of the appearance and disappearance data.
LOW and HIG refer, respectively, to MiniBooNE data with and without the three low-energy bins
from
$200 \, \text{MeV}$
to
$475 \, \text{MeV}$.
GAL refers to Gallium radioactive source experiment data analyzed according to Ref.~\cite{1006.3244}.
}
\end{center}
\end{table*}

\section{Global 3+1 Fits}
\label{Global}

In this section we present the results of the update of our analysis in Ref.~\cite{1107.1452}
which takes into account the new MiniBooNE antineutrino data discussed in Section~\ref{MiniBooNE}
and
the bound on $|U_{\mu4}|^2$ obtained from MINOS neutral-current data in Section~\ref{MINOS}.
As in Ref.~\cite{1107.1452},
we consider also the short-baseline
$\boss{\nu}{\mu}\to\boss{\nu}{e}$
data
of the
LSND \cite{hep-ex/0104049},
KARMEN \cite{hep-ex/0203021},
NOMAD \cite{hep-ex/0306037}
and 
MiniBooNE neutrino \cite{0812.2243}
experiments,
the short-baseline $\bar\nu_{e}$ disappearance data
of the
Bugey-3 \cite{Declais:1995su},
Bugey-4 \cite{Declais:1994ma},
ROVNO91 \cite{Kuvshinnikov:1990ry},
Gosgen \cite{Zacek:1986cu},
ILL \cite{Hoummada:1995zz}
and
Krasnoyarsk \cite{Vidyakin:1990iz}
reactor antineutrino experiments,
taking into account the new calculation of the reactor $\bar\nu_{e}$ flux
\cite{1101.2663,1106.0687}
which indicates a small $\bar\nu_{e}$ disappearance
(the reactor antineutrino anomaly \cite{1101.2755}),
the KamLAND \cite{0801.4589} bound on $|U_{e4}|^2$
(see Ref.~\cite{1012.0267}),
the short-baseline $\nu_{\mu}$ disappearance data
of the CDHSW experiment
\cite{Dydak:1984zq}
and
the constraints on $|U_{\mu4}|^2$ obtained in Ref.~\cite{0705.0107}
from the analysis of
the data of
atmospheric neutrino oscillation experiments.
We present global analyses of all these data without and with the
data of Gallium radioactive source experiments
(GALLEX
\cite{Anselmann:1995ar,Hampel:1998fc,1001.2731}
and
SAGE
\cite{Abdurashitov:1996dp,hep-ph/9803418,nucl-ex/0512041,0901.2200})
which indicate a $\nu_{e}$ disappearance
(the Gallium neutrino anomaly
\cite{hep-ph/9411414,Laveder:2007zz,hep-ph/0610352,0707.4593,0711.4222,0902.1992,1005.4599,1006.2103,1006.3244,1101.2755}).
We analyze the Gallium data according to Ref.~\cite{1006.3244}.

For the new MiniBooNE antineutrino data discussed in Section~\ref{MiniBooNE}
we consider two cases:
\begin{description}

\item[LOW]
All MiniBooNE neutrino and antineutrino data,
including the low-energy bins
from
$200 \, \text{MeV}$
to
$475 \, \text{MeV}$.

\item[HIG]
Only MiniBooNE neutrino and antineutrino data
with energy $E>475 \, \text{MeV}$.

\end{description}

\begin{figure*}[t!]
\begin{center}
\begin{tabular}{c}
\includegraphics*[width=0.7\linewidth]{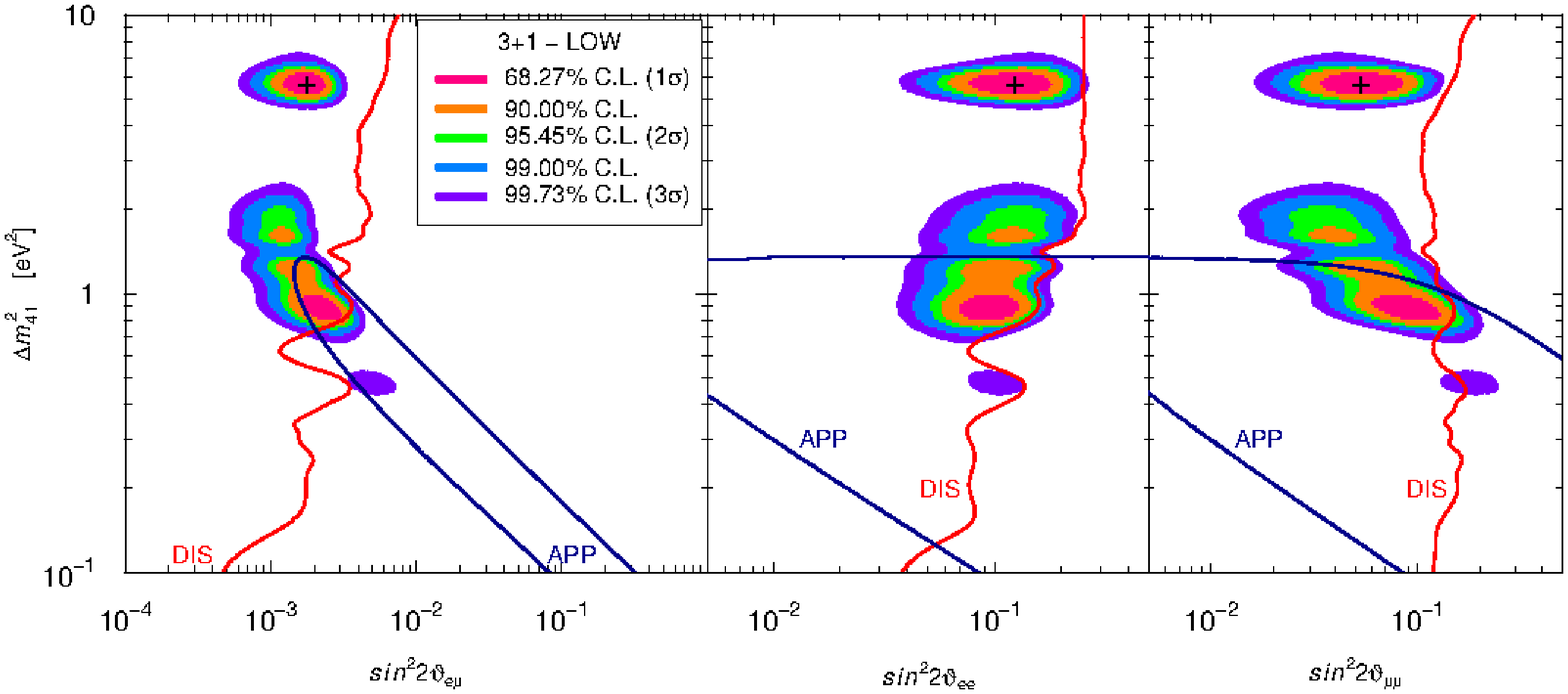}
\\
\includegraphics*[width=0.7\linewidth]{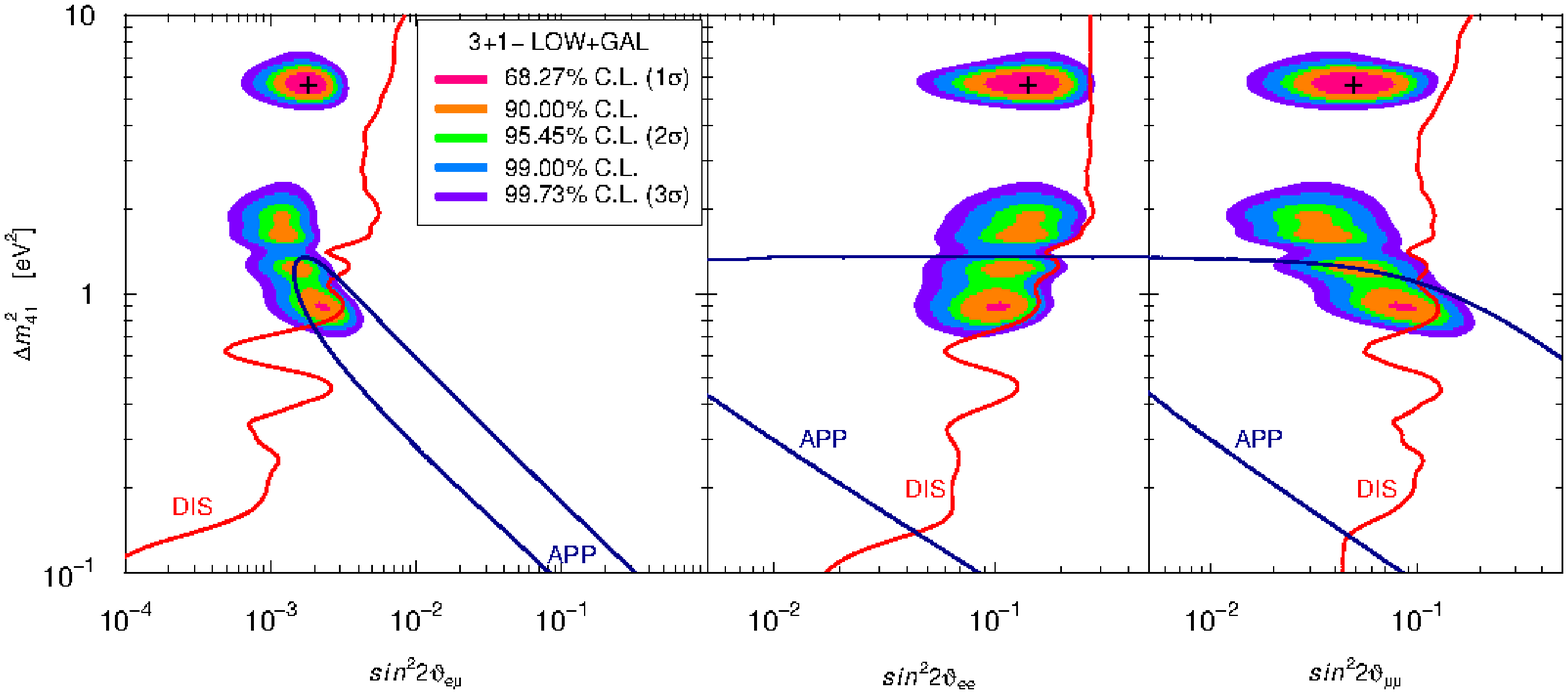}
\\
\includegraphics*[width=0.7\linewidth]{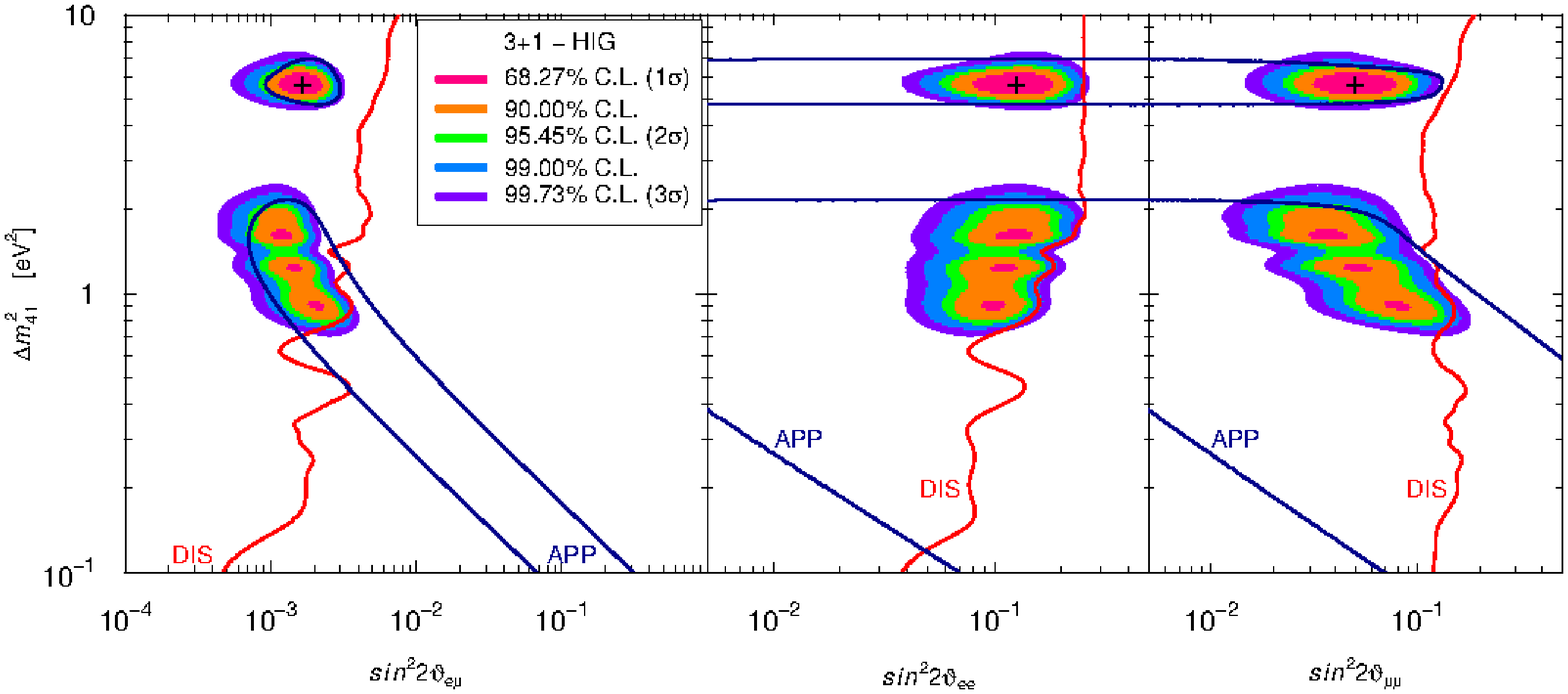}
\\
\includegraphics*[width=0.7\linewidth]{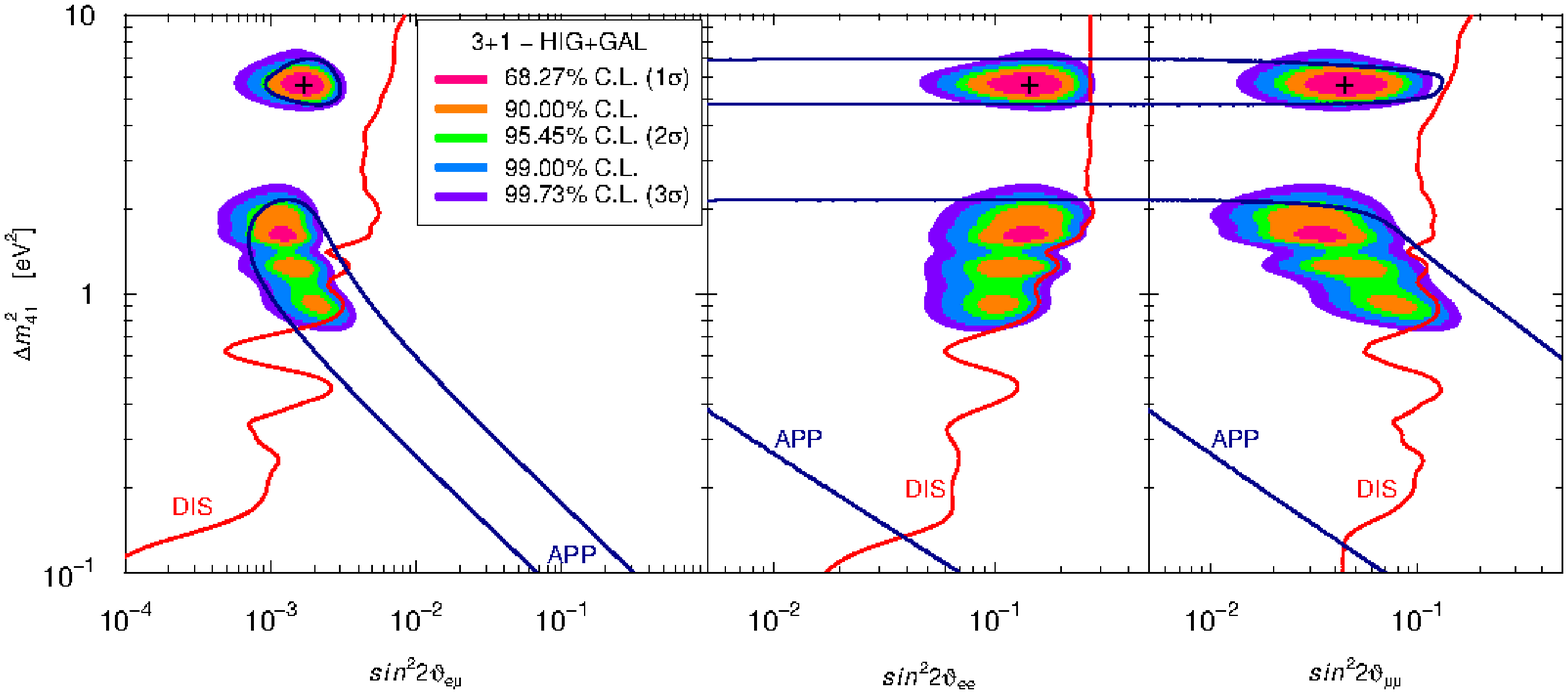}
\end{tabular}
\end{center}
\caption{ \label{con}
Allowed regions in the
$\sin^{2}2\vartheta_{e\mu}$--$\Delta{m}^2_{41}$,
$\sin^{2}2\vartheta_{ee}$--$\Delta{m}^2_{41}$ and
$\sin^{2}2\vartheta_{\mu\mu}$--$\Delta{m}^2_{41}$
planes in the four cases listed in Table~\ref{bef}.
The best-fit points are indicated by crosses.
The thick solid blue lines with the label APP
show the $3\sigma$ allowed regions obtained from the analysis of
$\protect\boss{\nu}{\mu}\to\protect\boss{\nu}{e}$
appearance data.
The thick solid red lines with the label DIS
show the $3\sigma$ allowed regions obtained from the analysis of
disappearance data.
}
\end{figure*}

Table~\ref{bef}
show the results of our global analyses for these two cases
without and with Gallium data (GAL).
The corresponding
allowed regions in the
$\sin^{2}2\vartheta_{e\mu}$--$\Delta{m}^2_{41}$,
$\sin^{2}2\vartheta_{ee}$--$\Delta{m}^2_{41}$ and
$\sin^{2}2\vartheta_{\mu\mu}$--$\Delta{m}^2_{41}$
planes are shown in Fig.~\ref{con}.

From Tab.~\ref{bef}
one can see that in all cases the
fit of the data without oscillations is rather bad
and the oscillations in the framework of
3+1 mixing have a good goodness of fit.
The goodness of fit is larger in the HIG case
than in the LOW case,
because the best fit of MiniBooNE data does not fit well
the low-energy bins,
as we have seen for the antineutrino data in Section~\ref{MiniBooNE}.
Gallium data slightly worsen the goodness of fit
because the best fit of the analysis of Gallium data \cite{1006.3244}
requires a value of $|U_{e4}|^2$
which is larger than the bound given by reactor antineutrino data
(see Ref.~\cite{1107.1452}).

The best-fit value of the 3+1 oscillation parameters
is rather stable under variations of the considered data sets
It points to a rather large value of
$\Delta{m}^2_{41}$,
which in a hierarchical scheme corresponds to
$m_{4}\approx2.4\,\text{eV}$.
This value is in tension with the limits given by
standard $\Lambda$CDM analyses of cosmological data
\cite{1006.5276,1102.4774,1104.0704,1104.2333,1106.5052,1108.4136}.
If it will be confirmed by future data,
it may indicate the existence of non-standard effects in the evolution of the Universe
\cite{1108.4136}.
However,
as one can see from
Fig.~\ref{con}
there are allowed regions
at
$\Delta{m}^2_{41}\approx1\,\text{eV}^2$
which are more compatible
with standard $\Lambda$CDM cosmology.
For example,
in the HIG fit there are three regions within $1\sigma$
at:
A)
$\Delta{m}^2_{41} \approx 1.6\,\text{eV}^2$,
$\sin^{2}2\vartheta_{e\mu} \approx 0.0012$,
$\sin^{2}2\vartheta_{ee} \approx 0.12$,
$\sin^{2}2\vartheta_{\mu\mu} \approx 0.037$;
B)
$\Delta{m}^2_{41} \approx 1.2\,\text{eV}^2$,
$\sin^{2}2\vartheta_{e\mu} \approx 0.0014$,
$\sin^{2}2\vartheta_{ee} \approx 0.11$,
$\sin^{2}2\vartheta_{\mu\mu} \approx 0.051$;
C)
$\Delta{m}^2_{41} \approx 0.91\,\text{eV}^2$,
$\sin^{2}2\vartheta_{e\mu} \approx 0.0020$,
$\sin^{2}2\vartheta_{ee} \approx 0.10$,
$\sin^{2}2\vartheta_{\mu\mu} \approx 0.078$.

Figure~\ref{con}
shows also the $3\sigma$ contours of the regions
allowed by appearance (APP) and disappearance (DIS)
data.
One can see that they are compatible with the global allowed regions.
It is interesting to note that the
values of
$\sin^{2}2\vartheta_{ee}$
and
$\sin^{2}2\vartheta_{\mu\mu}$
allowed by the analysis of appearance data
can be much smaller than those allowed by the global analysis.
The reason is that small values of
$\sin^{2}2\vartheta_{ee}$
and
$\sin^{2}2\vartheta_{\mu\mu}$
can be obtained not only with small values of
$|U_{e4}|^2$
and
$|U_{\mu4}|^2$,
respectively,
but also with
$|U_{e4}|^2$
and
$|U_{\mu4}|^2$
close to unity
(see Eq.~(\ref{survisin})).
Since one can fit the appearance data
with
$|U_{e4}|^2\simeq1$
and
$
\sin^{2} 2\vartheta_{e\mu}
\simeq
4 |U_{\mu4}|^2
$
or
$|U_{\mu4}|^2\simeq1$
and
$
\sin^{2} 2\vartheta_{e\mu}
\simeq
4 |U_{e4}|^2
$,
small values of
$\sin^{2}2\vartheta_{ee}$
and
$\sin^{2}2\vartheta_{\mu\mu}$
are allowed.
On the other hand,
in the global analysis
values of
$|U_{e4}|^2$
and
$|U_{\mu4}|^2$
close to unity
are forbidden,
respectively,
by the observation of solar
and atmospheric neutrino oscillations.
In our analysis we use the data of the
very-long-baseline KamLAND reactor antineutrino experiment
\cite{0801.4589}
which measured a disappearance of $\bar\nu_{e}$
due to
$\Delta{m}^2_{21}$
which is compatible with solar neutrino oscillations.
The KamLAND measurements require a relatively large value of
$|U_{e1}|^2+|U_{e2}|^2$,
which by unitarity
constrain $|U_{e4}|^2$ to be much smaller than unity
(see Ref.~\cite{1012.0267}).
In a similar way,
the observation of atmospheric $\boss{\nu}{\mu}$ oscillations
due to
$\Delta{m}^2_{31}$
requires a relatively large values of
$|U_{\mu1}|^2+|U_{\mu2}|^2+|U_{\mu3}|^2$,
which by unitarity
constrain $|U_{\mu4}|^2$ to be much smaller than unity.
In our analysis we use the bound on
$|U_{\mu4}|^2$
obtained in the analysis
of atmospheric neutrino data
presented in Ref.~\cite{0705.0107}.

In Fig.~\ref{con}
one can see that for $\Delta{m}^2_{41}\gtrsim1\,\text{eV}^2$
large values of
$\sin^{2}2\vartheta_{\mu\mu}$
are not allowed by the analysis of appearance data.
The reason is that,
as explained in Section~\ref{MiniBooNE},
we fit the MiniBooNE
$\boss{\nu}{\mu}\to\boss{\nu}{e}$
data together with the MiniBooNE
$\boss{\nu}{\mu}\to\boss{\nu}{\mu}$
data which have correlated uncertainties.
The MiniBooNE
$\boss{\nu}{\mu}\to\boss{\nu}{\mu}$
data constrain the disappearance of
$\boss{\nu}{\mu}$'s
for $\Delta{m}^2_{41}\gtrsim1\,\text{eV}^2$
\cite{0903.2465},
limiting the allowed value of
$\sin^{2}2\vartheta_{\mu\mu}$.

The last line of Tab.~\ref{bef}
gives the parameter goodness-of-fit \cite{hep-ph/0304176}
in the four analyses,
which has been obtained by comparing the global best-fit with the sum of the
best fits of the appearance and disappearance data.
We think that this parameter goodness-of-fit is more reliable than the
parameter goodness-of-fit obtained by comparing data in favor and against
short-baseline neutrino oscillations,
which has been considered in several previous analyses,
including ours
\cite{1107.1452}.
The reason is that the distinction between appearance and disappearance data
is made a priori,
without considering the data.
In this case
$
\Delta\chi^2_{\text{min}}
=
(\chi^2_{\text{min}})_{\text{app+dis}}
-
(\chi^2_{\text{min}})_{\text{app}}
-
(\chi^2_{\text{min}})_{\text{dis}}
$
is a random variable with a $\chi^2$ distribution with
2 degrees of freedom,
as shown in Ref.~\cite{hep-ph/0304176}.
On the other hand,
comparing data in favor and against
short-baseline neutrino oscillations
one chooses the two sets of data in order to obtain always the worse
$\Delta\chi^2_{\text{min}}$
allowed by the data.
In this case $\Delta\chi^2_{\text{min}}$
does not have a $\chi^2$ distribution
and the parameter goodness-of-fit is underestimated by assuming
a $\chi^2$ distribution with
2 degrees of freedom.
Moreover,
after the discovery of the reactor antineutrino anomaly
\cite{1101.2755}
it is not clear
if the reactor antineutrino data should be put in the group
of experiments in which there is no evidence of
short-baseline neutrino oscillations,
as traditionally done,
or in the group
of experiments in favor of
short-baseline neutrino oscillations.
Therefore,
we advocate the robust
parameter goodness-of-fit \cite{hep-ph/0304176}
obtained from appearance and disappearance data.

The values of such parameter goodness-of-fit
listed
in the last line of Tab.~\ref{bef}
show that there is a tension between appearance and disappearance data.
This tension is severe in the two LOW fits.
The reason can be understood by looking at the two corresponding panels
in Fig.~\ref{con}.
One can see that
for appearance data there is no $3\sigma$ allowed region
around the best-fit point,
because the low-energy MiniBooNE data are not fitted well
by such high values of $\Delta{m}^2_{41}$ and small mixing.
Hence the best fit point lies out of the
region of overlap of the $3\sigma$ allowed regions
of appearance and disappearance data.
This is a symptom of a severe tension.

The tension between appearance and disappearance data
is reduced in the HIG fits,
for which there is a $3\sigma$ allowed region
around the best-fit point,
as one can see from the corresponding panels
in Fig.~\ref{con}.
The resulting parameter goodness-of-fit,
about 1\%,
is not large, but also not small enough to reject the fit
with reasonable confidence.

Table~\ref{bef} and Fig.~\ref{con}
show that the Gallium data do not have a large impact on the
results of the fit.
The reason is that the data points are only four,
much less than the reactor data points which give information on
the same probability of $\nu_{e}$ and $\bar\nu_{e}$ disappearance.
The main effect of Gallium data is to favor values of
$\Delta{m}^2_{41}$ larger than about $1 \, \text{eV}^2$.
Indeed,
in the HIG+GAL panel in Fig.~\ref{con}
there are no regions allowed at $1\sigma$
below about $1.3 \, \text{eV}^2$.

In comparison with the allowed regions
of the 3+1 oscillation parameters
obtained in Ref.~\cite{1107.1452},
the inclusion in the analysis of the MINOS data discussed in Section~\ref{MINOS}
has the effect of disfavoring the regions
with
$\Delta{m}^2_{41} \lesssim 1 \, \text{eV}^2$
and moving the best-fit value of
$\Delta{m}^2_{41}$
from the value of $0.9\,\text{eV}^2$ obtained in Ref.~\cite{1107.1452}
to $\Delta{m}^2_{41} \approx 5.6\,\text{eV}^2$.

\section{Conclusions}
\label{Conclusions}

The results of
our analysis of short-baseline neutrino oscillation data show that
the data can be fitted in the framework of 3+1 neutrino mixing,
which requires the existence of a sterile neutrino with mass
at the eV scale.

The fit has a tension due to the lack of observation of
enough
short-baseline disappearance of $\bar\nu_{e}$ and $\nu_{\mu}$
to fully explain the
$\bar\nu_{\mu}\to\bar\nu_{e}$
signal observed in the LSND and MiniBooNE experiments.
However,
we found that the appearance and disappearance data
are marginally compatible
if we neglect the data on the MiniBooNE low-energy anomaly,
which may have an explanation different from
$\boss{\nu}{\mu}\to\boss{\nu}{e}$
oscillations.
In any case,
we think that the neutrino oscillation explanation of the data
cannot be dismissed with a light heart,
because besides the LSND and MiniBooNE indications
in favor of a short-baseline
$\bar\nu_{\mu}\to\bar\nu_{e}$
signal
we have
the reactor antineutrino anomaly
\cite{1101.2755}
and the Gallium neutrino anomaly
\cite{1006.3244}
in favor,
respectively,
of short-baseline
$\bar\nu_{e}$ and $\nu_{e}$
disappearance
which could be due to the same
squared-mass difference.

Since the recent MiniBooNE antineutrino data
\protect\cite{1111.1375,Djurcic-NUFACT2011}
do not show a large difference from the neutrino data
\cite{0812.2243},
there is no serious motivation to consider the more complicated
3+2 neutrino mixing,
which would allow for a possible CP-violating difference
between
neutrino and antineutrino transitions
\cite{hep-ph/0305255,hep-ph/0609177,0906.1997,0705.0107,1007.4171,1103.4570,1107.1452}.
Moreover,
as we have shown in Ref.~\cite{1107.1452},
3+2 mixing cannot resolve the tension between appearance and disappearance data.
Finally,
the hierarchical 3+1 scheme (see Eq.~(\ref{hierarchy}))
is favored over a hierarchical 3+2 scheme
by standard $\Lambda$CDM analyses of cosmological data
\cite{1006.5276,1102.4774,1104.0704,1104.2333,1106.5052,1108.4136},
which disfavor sums of neutrino masses much larger than 1 eV,
and
by Big-Bang Nucleosynthesis data,
which allow the existence of one sterile neutrino
\cite{astro-ph/0408033,1001.4440},
but not more
\cite{1103.1261,1108.4136}
(keeping however in mind the caveat
that these bounds refer to the number of sterile neutrinos which
have been fully thermalized in the early Universe).

Hence,
in this paper we considered only
3+1 neutrino mixing,
which is attractive
for the natural correspondence of the existence
of one new entity (a sterile neutrino) with
the observation of a new effect (short-baseline oscillations).

The results of our fit excluding the MiniBooNE low-energy anomaly
lead to a best fit at
$\Delta{m}^2_{41} \approx 5.6\,\text{eV}^2$,
which is larger than that
obtained in Ref.~\cite{1107.1452}
(about $0.9\,\text{eV}^2$)
because of the new
MINOS constraints discussed in Section~\ref{MINOS}.
The new best fit value of
$\Delta{m}^2_{41}$
is rather large
in comparison with the standard cosmological bounds
\cite{1006.5276,1102.4774,1104.0704,1104.2333,1106.5052,1108.4136}
and may indicate the existence
of non-standard effects in the evolution of the Universe
\cite{1108.4136}.
However,
there are three regions within $1\sigma$
at
$
\Delta{m}^2_{41}
\approx
1.6
\,,\,
1.2
\,,\,
0.91
\,
\text{eV}^2
$
which may be compatible with
the standard cosmological bounds.

We have also shown
that the data on the Gallium neutrino anomaly
favor values of
$\Delta{m}^2_{41}$ larger than about $1 \, \text{eV}^2$,
but their impact is small because the results of the analysis are
dominated by the more abundant reactor antineutrino data,
which give information on
the same probability of $\nu_{e}$ and $\bar\nu_{e}$ disappearance.

\bigskip
\centerline{\textbf{Acknowledgments}}
\medskip

We would like to thank
Z.~Djurcic,
W.~Louis,
G.~Mention,
T.~Schwetz,
A.~Smirnov,
A.~Sousa,
and
O.~Yasuda
for useful discussions.

\bibliography{bibtex/nu}

\end{document}